\documentclass[shortnames]{article}
\usepackage{common}

\begin{document}

\author{Sean Martin\\UMBC\\ \and	
Andrew~M. Raim\\U.S.~Census Bureau\\ \and
Wen Huang \\Rice University\\ \and	
Kofi~P. Adragni\footnote{kofi@umbc.edu} \\UMBC}

\title{ManifoldOptim: An R Interface to the ROPTLIB Library for Riemannian Manifold Optimization}

\maketitle

\begin{abstract}
Manifold optimization appears in a wide variety of computational problems in the applied sciences. In recent statistical methodologies such as sufficient dimension reduction and regression envelopes, estimation relies on the optimization of likelihood functions over spaces of matrices such as the Stiefel or Grassmann manifolds. Recently, \citet*{HAG2016} have introduced the library {ROPTLIB}, which provides a framework and state of the art algorithms to optimize real-valued objective functions over commonly used matrix-valued Riemannian manifolds. This article presents {ManifoldOptim}, an {R} package that wraps the {C++} library {ROPTLIB}. {ManifoldOptim} enables users to access functionality in {ROPTLIB} through {R} so that optimization problems can easily be constructed, solved, and integrated into larger {R} codes. Computationally intensive problems can be programmed with {Rcpp} and {RcppArmadillo}, and otherwise accessed through {R}. We illustrate the practical use of {ManifoldOptim} through several motivating examples involving dimension reduction and envelope methods in regression.
\end{abstract}

\blfootnote{\noindent Disclaimer: This article is released to inform interested parties of ongoing research and to encourage discussion of work in progress. Any views expressed are those of the authors and not necessarily those of the U.S.~Census Bureau.}

\section{Introduction}
\label{sec:intro}

This article presents {ManifoldOptim} \citep{ManifoldOptim2016}, an {R} \citep{Rco} package for the optimization of real-valued functions on Riemannian manifolds. {ManifoldOptim} is a wrapper for {ROPTLIB}, a {C++} optimization library by \citet{HAG2016}, which provides a variety of cutting edge algorithms for optimization on manifolds and a framework for constructing such optimization problems.

There is a growing literature on manifolds and manifold optimization. Manifolds are generalizations of smooth curves and surfaces within higher dimensions. Riemannian manifolds are a class of manifolds which have rules to compute distances and angles. Some manifolds are especially abundant in the literature because of their wide applicability; for example, the differential geometry of Grassmann manifolds has been well studied \citep{won, e+a+s, chi, AMS08}. Mathematical introductions to manifolds can be found in \citet{leej00, leej03} and \citet{tl11}.

Optimization on Riemannian manifolds generalizes Euclidean optimization algorithms to curved and smoothed surfaces \citep{AMS08, won, e+a+s, chi}. Applications are found in many areas including signal processing \citep{c+g}, data mining \citep{b+d+j}, and control theory \citep{p+l+d}. Many eigenvalue problems are, in fact, optimization problems on manifolds \citep{AMS08, e+a+s}. Recent developments in likelihood-based sufficient dimension reduction \citep{c+f+08,c+l, c+f+09} and regression envelope models \citep{c+l+c, SuC11,SuC12,CookS13, CookZ15} rely on optimizations over manifolds.

Formulating optimization algorithms on manifolds requires endowing the manifolds with a differentiable structure. The gradient and Hessian are often used in Euclidean optimization to find the direction for the next iterate, but these notions must be extended to honor the curved structure of the manifold within Euclidean space. Each manifold is unique with respect to its differential geometry and requires some individual consideration. In this article, we focus on the following manifolds: Euclidean, Grassmann, low-rank manifold, manifold of $n$-dimensional vectors on the unit sphere, Stiefel, and space of positive definite matrices. We also consider spaces formed by taking Cartesian products of these manifolds. Available optimization algorithms include the: Riemannian trust-region Newton \citep{ABG07}, Riemannian symmetric rank-one trust-region \citep[RTRSR1;][]{hag15}, limited-memory \citep[RTRSR1;][]{hag15}, Riemannian trust-region steepest descent \citep{AMS08}, Riemannian line-search Newton \citep{AMS08}, Riemannian Broyden family \citep{HGA15}, Riemannian Broyden-Fletcher-Goldfarb-Shannon \citep[RBFGS;][]{RW12, HGA15}, limited memory \citep[RBFGS;][]{HGA15}, Riemannian conjugate gradients \citep{n+w, AMS08, si13}, and Riemannian steepest descent \citep{AMS08}.

Consider the problem of minimizing $f(\Urm)$ for $\Urm \in \Mcal$ where $\Mcal$ is a Riemannian manifold. For example, let $\Urm$ be a $p \times d$ semi-orthogonal matrix where $\Urm^T\Urm=\Irm_d$ and $d < p$. This problem naturally lies in a Stiefel manifold, where orthonormality of the argument $\Urm$ is intrinsic to the manifold. When $f$ is invariant under right orthogonal transformations of $\Urm$, so that $f(\Urm)=f(\Urm \Orm)$ for any $d \times d$ orthogonal matrix $\Orm$, then the problem lies in a Grassmann manifold. If $\Urm$ is a $p \times p$ symmetric positive definite (SPD) matrix, such as a covariance matrix, the problem is naturally defined on the manifold of SPD matrices. Methods abound in the literature for optimization problems with linear constraints \citep{fle, n+w}. Constrained problems sometimes become unconstrained by an appropriate transformation. The constraints in the above examples are naturally honored by restricting to the respective manifolds.

In a typical use of {ManifoldOptim}, the user provides an objective function, its gradient and Hessian, and a specification of the manifold, the solver, and the optimization method to be used. Numerical gradient and Hessian functions are provided, as defaults, for problems where closed-form expressions are not easily programmed. {ManifoldOptim} users can construct problems in plain {R} for convenience and quick prototyping. If performance becomes a concern, users can implement the objective, gradient, and Hessian functions in {C++} and otherwise interact with {ManifoldOptim} through {R}. We make use of the {Rcpp} \citep{EddelbuettelFrancois2011} and {RcppArmadillo} \citep{EddelbuettelSanderson2014} packages to reduce the burden of {C++} programming for {R} users, and to facilitate seamless integration with {R}. The {ROPTLIB} library itself is written in {C++} and uses standard linear algebra libraries such as {BLAS} and {LAPACK} \citep{laug} to ensure generally good performance.

{ROPTLIB} can be readily used in {Matlab} through its included \texttt{mex}-based wrapper. Interfaces to {ROPTLIB} for other high level languages, such as {Julia}, have also been developed \citep{HAG2016}. Other manifold optimization software packages are found in the literature: {sg\_min} written by \citet{lip} which was adapted from \citet{e+a+s}, and {Manopt} of \citet{manopt} are available in {Matlab}. In {R}, \citet{AdragniCW10} developed {GrassmannOptim} specifically for the Grassmann manifold. To our knowledge, there are no other publicly available {R} packages for manifold optimization.

The following notations are adopted throughout this manuscript: $\Gcal_{d,p}$ represents the Grassmann manifold, the set of all $d$-dimensional subspaces in $\Rbb^{p}$; $\Scal_{d,p}$ represents the Stiefel manifold, the set of all $d$-dimensional orthonormal matrices in $\Rbb^p$; $\Scal^{+}_{p}$ represents the manifold of all $p \times p$ symmetric positive definite matrices, and $\Scal^p$ is the unit sphere manifold.

The remainder of this article is organized as follows. Section \ref{sec:alg} gives a brief description of the algorithms that guided the programming of {ManifoldOptim}. Section~\ref{sec:usage} demonstrates usage of the package through brief examples. Section \ref{sec:examples} presents several applications where manifold optimization is useful in statistics. {R} code for these applications is somewhat lengthy and therefore is provided as supplementary material. Discussions and conclusions follow in section~\ref{sec:conclusions}. The {ManifoldOptim} package is available from the Comprehensive R Archive Network at \url{https://CRAN.R-project.org/package=ManifoldOptim}.

\section{Optimization on Manifolds}
\label{sec:alg}

Manifold optimization appears in a wide variety of computational problems in the applied sciences. It has recently gained prominence in the statistics literature. Optimization over a Grassmann manifold is used in several models for sufficient dimension reduction (SDR) in regression \citep{Cook98, Cook07}, including covariance reducing models \citep{c+f+08} and likelihood acquired directions \citep{c+f+09}. \citet{AdragniCW10} provide some relevant examples. Essentially, a manifold optimization is characterized by symmetry or invariance properties of the objective function.

Consider the Brockett problem \citep[section 4]{AMS08}, which amounts to minimizing the objective function
\begin{equation} \label{eq:brock}
f(X) = \tr(X^TBXD), \quad X \in \Rbb^{p \times d}
\end{equation}
subject to $X^TX=I_p$. Here, $D = \diag(\mu_1, \dots, \mu_p)$ with $\mu_1 \geq \dots \geq \mu_p \geq 0$ and $B \in \Rbb^{n \times n}$ is a given symmetric matrix. The optimization can be carried out over a Stiefel manifold $\Scal_{d,p}$.
It is known that a global minimizer of $f$ is the matrix whose columns are the eigenvectors of $B$ corresponding to the $d$ smallest eigenvalues $\mu_{p-d+1}, \dots, \mu_p$. This is essentially an eigenvalue problem reminiscent of the generalized Rayleigh quotient for discriminant analysis \citep{AdragniCW10}. It will be later used in section~\ref{sec:usage} to illustrate practical use of the package.

We next present three algorithms in statistics that can be formulated as manifold optimization problems. The first is the recently developed Minimum Average Deviance Estimation method for SDR \citep{AdragniRA16}. Second is Cook's Principal Fitted Components model for SDR \citep{Cook07}. Third is an envelope method for regression initially proposed by \citet{c+l+c}. These problems will be explored further in section~\ref{sec:examples}.

\subsection{Minimum Average Deviance Estimation}%
\label{sec:made}
Minimum Average Deviance Estimation (MADE) is a dimension reduction method based on local regression \citep{AdragniRA16}. This approach was first proposed by \citet{Xia02} under the assumption of additive errors. Suppose $Y \in \Rbb$ is a response, $X$ is a $p$-dimensional predictor, and the distribution of $Y \mid X$ is given by an exponential family distribution of the form
\begin{equation}
f(Y \mid \theta (X) ) = f_0(Y, \phi) \exp\left\{ \frac{Y \cdot \theta(X) - b(\theta(X))}{a(\phi)} \right\}
\label{eqn:exp-fam}
\end{equation}
with dispersion parameter $\phi$. The canonical parameter $\theta(X)$ possesses the main information that connects $Y$ to $X$; it relates to the mean function $E(Y \mid X)$ through a link function $g$ so that $g(E(Y \mid X))= \theta(X)$. Let $(Y_i, X_i)$, $i=1, \ldots, n$, represent an independent sample from the distribution of $(Y, X)$ so that $Y_i \mid X_i$ has the distribution \eqref{eqn:exp-fam}. If there exists a semi-orthogonal $B \in \Rbb^{p \times d}$ with $d < p$ and $\theta(X) = \vartheta(B^TX)$ for some function $\vartheta$, then $X \in \Rbb^p$ can be replaced by $B^T X \in \Rbb^d$ in the regression of $Y$ on $X$, and the subspace $\Scal_B$ is a dimension reduction subspace.

Regression based on the local log-likelihood evaluated at a given $X_0 \in \Rbb^p$ can be written as
\begin{equation*}
L_{X_0}(\alpha_0, \gamma_0, B) = \sum_{i=1}^n w(X_i, X_0) \log f(Y_i \mid \alpha_0 + \gamma_0^T B^T (X_i - X_0)),
\end{equation*}
using the first-order Taylor expansion
\begin{align*}
\vartheta(B^T X_i) &\approx \vartheta(B^T X_0) + [\nabla\vartheta(B^T X_0)]^T (B^T X_i - B^T X_0) \\
&= \alpha_0 + \gamma_0^T B^T (X_i - X_0),
\end{align*}
taking $\alpha_0 = \vartheta(B^T X_0)$ and $\gamma_0 = \nabla\vartheta(B^T X_0)$. The $0$ subscript in $\alpha_0, \gamma_0$ denotes that the parameters vary with the choice of $X_0$. The weights $w(X_1, X_0), \ldots, w(X_n, X_0)$ represent the contribution of each observation toward $L_{X_0}(\alpha_0, \gamma_0, B)$. A local deviance for the $j$th observation can be defined as
\begin{align}
D(Y_j, \vartheta(B^T X_j)) = 2 \left[ \max_{\vartheta}\log f(Y_j \mid \vartheta) - L_{X_j}(\alpha_j, \gamma_j, B) \right].
\end{align}
The term $\textstyle{\max_{\vartheta}\log f(Y_j \mid \vartheta)}$ is the maximum likelihood achievable for an individual observation. Consider minimizing the average deviance $n^{-1} \textstyle{\sum_{j=1}^{n}} D(Y_j, \vartheta(B^T X_j))$ with respect to $(\alpha_j, \gamma_j) \in \Rbb^{d+1}$ for $j = 1, \ldots, n$ and $B \in \Rbb^{p \times d}$ such that $B^T B = I$. This is equivalent to maximizing
\begin{equation}\label{eqn:made-glm-obj}
Q(\alphabf, \gammabf, B) = \sum_{j=1}^n \sum_{i=1}^n w(X_i, X_j)
\left\{Y_i (\alpha_j + \gamma_j^T B^T (X_i - X_j)) - b(\alpha_j + \gamma_j^T B^T (X_i - X_j))
\right\}
\end{equation}
over the same space.
The kernel weight function is taken to be
\begin{align*}
w(X_i, X_0) = \frac{K_{\Hrm}(X_i - X_0) }{ \textstyle{\sum_{j=1}^n} K_{\Hrm}(X_j - X_0) },
\quad
K_{\Hrm}(\urm)=|\Hrm|^{-1}K(\Hrm^{-1/2}\urm),
\end{align*}
where $K(u)$ denotes one of the usual multidimensional kernel density functions and the bandwidth $\Hrm$ is a $p \times p$ symmetric and positive definite matrix. We may also consider refined weights $w(B^T X_i, B^T X_0)$ which make use of the unknown $B$.

For any orthogonal matrix $O \in \Rbb^{d \times d}$, $\gamma^T B^T = \gamma^T O O^T B^T$, which implies that $\gamma$ and $B$ are not uniquely determined but obtained up to an orthogonal transformation. Furthermore, refined weights based on the Gaussian kernel with $H = hI$ depend on $B$ only through $B B^T = B O O^T B^T$. In this setting, the MADE problem is invariant to orthogonal transformation of $B$ in the sense that
\begin{equation*}
Q(\alphabf, \gammabf, B) = Q(\alphabf, O^T \gamma_1, \ldots, O^T \gamma_n, B O).
\end{equation*}
The joint parameter space of $(\alphabf, \gammabf, B)$ is a product manifold $\Rbb^{n (d+1)} \times \Gcal_{d, p}$. However, in their initial work, \citet{AdragniRA16} used an iterative algorithm to maximize the objective function \eqref{eqn:made-glm-obj}:
\begin{enumerate}
\item Fix $B$ and maximize over $(\alpha_j, \gamma_j) \in \Rbb^{d + 1}$ using Newton-Raphson for $j = 1, \ldots, n$.
\item Fix $\alphabf$ and $\gammabf$ and the weights $w_{ij}$, and maximize over $B \in \Scal_{d,p}$.
\item Update the weights $w_{ij} = w(B^T X_i, B^T X_j)$ if refined weights are desired.
\end{enumerate}
These iterations are started from an initial estimate for $B$, and repeated until changes in $B$ are smaller than some specified threshold.

\subsection{Principal Fitted Components}
\label{sec:pfc}
The principal fitted components (PFC) model was initially proposed by \citet{Cook07} as a likelihood-based method for sufficient dimension reduction in regression.
Let $\Xrm$ be a $p$-vector of random predictors and $Y$ be the response. Using the stochastic nature of the predictors, \citet{Cook07} proposed the model
\begin{equation}
\Xrm_y = \mu + \Gamma \beta \frm_y + \Delta^{1/2}\varepsilon. \label{pfcmodel}
\end{equation}
Here, $\Xrm_y$ denotes the conditional $\Xrm$ given $Y=y$, $\frm_y \in \Rbb^r$ is a user-selected function of the response, which helps capture the dependency of $\Xrm$ on $Y$. The other parameters are $\mu \in \Rbb^p$, $\Gamma \in \Rbb^{p \times d}$ is a semi-orthogonal matrix and $\beta \in \Rbb^{d \times r}$. The error term $\varepsilon \in \Rbb^{p}$ is assumed to be normally distributed with mean 0 and variance $\Delta$.

This model can be seen as a way to partition the information in the predictors given the response into two parts: one part $\Gamma \beta \frm_y=\Erm(\Xrm_y -\Xrm)$ that is related to the response and another part $\mu + \varepsilon$ that is not. The form of the related part suggests that the translated conditional means $\Erm(\Xrm_y -\Xrm)$ fall in the $d$-dimensional subspace $\Scal_{\Gamma}$.

\citet{Cook07} showed that a sufficient reduction of $\Xrm$ is $\eta^T \Xrm$, where $\eta= \Delta^{-1}\Gamma$, so that $\Xrm$ can be replaced by $\eta^T\Xrm$ without loss of information about the regression of $Y$ on $\Xrm$. However, as $\eta^T \Xrm$ is a sufficient reduction, any $d \times d$ orthogonal matrix $\Orm$, $\Orm^T \eta^T \Xrm$ is also sufficient. Thus $\Gamma$ is not estimable but the subspace spanned by its columns is estimable. The goal of an analysis is then to estimate the subspace $\Scal_{\Gamma}$ spanned by the columns of $\Gamma$.

The estimation of $\Scal_{\Gamma}$ depends on the structure of $\Delta$. Various structures for $\Delta$ can be modeled, including isotropic $\Delta=\sigma^2 \Irm$, diagonal $\Delta=\diag(\sigma_1^2, ..., \sigma_p^2)$, the heterogeneous structure $\Delta=\Gamma \Omega \Gamma^T + \Gamma_0 \Omega_0 \Gamma_0^T$, and the general unstructured $\Delta > 0$ cases. In the heterogeneous structure of $\Delta$, $\Gamma_0$ is the orthogonal completion of $\Gamma$ such that $(\Gamma, \Gamma_0)$ is a $p \times p$ orthogonal matrix. The matrices $\Omega \in \Rbb^{d \times d}$ and $\Omega_0 \in \Rbb^{(p-d) \times (p-d)}$ are assumed to be symmetric and full-rank.

We now focus on maximum likelihood estimation in PFC under the heterogeneous and general unstructured forms of $\Delta$. Assuming that a set of $n$ data points is observed, let $\Xrmbar$ be the sample mean of $\Xrm$ and let $\Xbb$ denote the $n \times p$ matrix with rows $(\Xrm_y - \Xrmbar)^T$. Let $\Fbb$ denote the $n \times r$ matrix with rows $(\frm_y-\frmbar)^T$ and set $\Prm_{\Fbb}$ to denote the linear operator that projects onto the subspace spanned by the columns of $\Fbb$. Also let $\Xbbhat=\Prm_{\Fbb} \Xbb$ denote the $n \times p$ matrix of the fitted values from the multivariate linear regression of $\Xrm$ on $\frm_y$. Let $\Sigmahat=\Xbb^T\Xbb/n$ and $\Sigmahat_{\fit}=\Xbbhat^T \Xbbhat/n$. For the heterogeneous and the unstructured $\Delta$, the log-likelihood functions are respectively
\begin{eqnarray}
\Lcal(\Omega, \Omega_0, \Gamma) & = &  -\frac{n}{2} \log|\Omega| -\frac{n}{2}\log|\Omega_0| - \tr\left\{\Gamma^T\Sigmahat_{\res} \Gamma \Omega^{-1} - \Gamma_0^T\Sigmahat \Gamma_0 \Omega_0^{-1} \right\}, \label{env}\\
\Lcal(\Delta, \Gamma) & = &  -\frac{n}{2} \log|\Delta| - \frac{n}{2} \tr \left\{ \left[\Sigmahat - \Sigmahat_{\fit} \Delta^{-1} \Gamma(\Gamma^T \Delta^{-1} \Gamma)^{-1} \Gamma^T \right] \Delta^{-1} \right\}. \label{unst}
\end{eqnarray}
These functions are real-valued, with parameter spaces expressed as Cartesian products of manifolds $(\Omega, \Omega_0, \Gamma) \in \Scal_{d}^{+} \times \Scal_{p-d}^{+} \times \Gcal_{d,p}$ and $(\Delta, \Gamma) \in \Scal_p^{+} \times \Gcal_{d,p}$, respectively. Maximum likelihood estimators are obtained as
\begin{eqnarray}
(\Omegahat, \Omegahat_0, \Scalhat_{\Gamma}) & = & \argmax_{(\Omega, \Omega_0, \Scal_{\Gamma})} \Lcal(\Omega, \Omega_0, \Gamma), \label{mlepfc1} \\
(\Deltahat, \Scalhat_{\Gamma}) & = & \argmax_{(\Delta, \Scal_{\Gamma})} \Lcal(\Delta, \Gamma). \label{mlepfc2}
\end{eqnarray}
With multiple parameters constrained to different spaces, a common approach to maximum likelihood estimation is to optimize one parameter at a time while keeping the others fixed, cycling through such steps until meeting convergence criteria for the overall problem. For these situations, {ManifoldOptim} supports a product space of manifolds which are composed from simpler manifolds and can include Euclidean spaces. Here, optimization algorithms work over all parameters jointly rather than cycling through partial optimizations. In our experience, joint optimization has produced comparable or superior results to the cycling approach with a reduced programming burden.

\subsection{Envelope Method for Regression}
\label{sec:env}

Enveloping is a novel and nascent method initially introduced by \citet{c+l+c} that has the potential to produce substantial gains in efficiency for multivariate analysis. The initial development of the envelope method was in terms of the standard multivariate linear model
\begin{equation} \label{linmodel}
Y = \mu + \beta X + \varepsilon,
\end{equation}
where $\mu \in \Rbb^{r}$, the random response $Y \in \Rbb^{r}$, the fixed predictor vector $X \in \Rbb^{p}$ is centered to have sample mean zero, and the error vector $\varepsilon \sim N(0, \Sigma)$. In this context some linear combinations of $Y$ are immaterial to the regression because their distribution does not depend on $X$, while other linear combinations of $Y$ do depend on $X$ and are thus material to the regression. In effect, envelopes separate the material and immaterial parts of $Y$, and thereby allow for gains in efficiency.

Suppose that we can find a subspace $\Scal \in \Rbb^r$ so that
\begin{equation} \label{state}
Q_{\Scal}Y \mid X \sim  Q_{\Scal}Y\;\; \text{  and  }\;\; Q_{\Scal}Y \indep P_{\Scal}Y \mid X,
\end{equation}
where $\sim$ means identically distributed, $P_{\Scal}$ projects onto the subspace $\Scal$
and $Q_{\Scal} = I_r - P_{\Scal}$. For any $\Scal$ with those properties, $P_{\Scal}Y$ carries all of the material information and perhaps some immaterial information, while $Q_{\Scal}$ carries just immaterial information. Denoting
$\Bcal := \spn(\beta)$, expressions \eqref{state} hold if and only if $\Bcal \subseteq \Scal$ and $\Sigma=\Sigma_{\Scal}+ \Sigma_{\Scal^{\perp}}$, where $\Sigma_{\Scal}=\var(P_{\Scal}Y)$ and $\Sigma_{\Scal^{\perp}}=\var(Q_{\Scal}Y)$. The subspace $\Scal$ is not necessarily unique nor minimal, because there may be infinitely many subspaces that satisfy these relations in a particular problem. \citet{c+l+c} showed that $\Scal$ is a reducing subspace of $\Sigma$ if and only if $\Sigma= \Sigma_{\Scal} + \Sigma_{\Scal^{\perp}}$, and defined the minimal subspace to be the intersection of all reducing subspaces of $\Sigma$ that contain $\Bcal$, which is called the $\Sigma$-envelope of $\Bcal$ and denoted as $\Ecal_{\Sigma}(\Bcal)$. Let $u=\dim\{\Ecal_{\Sigma}(\Bcal)\}$. Then
$$\Bcal \subseteq \Ecal_{\Sigma}(\Bcal)\;\; \text{ and } \;\; \Sigma=\Sigma_{\Scal}+ \Sigma_{\Scal^{\perp}},$$

where $\Ecal_{\Sigma}(\Bcal)$ is shortened to $\Ecal$ for subscripts. These relationships establish a unique link between
the coefficient matrix $\beta$ and the covariance matrix $\Sigma$ of model \eqref{linmodel}, and it is this link that has the potential to produce gains in the efficiency of estimates of $\beta$ relative to the standard estimator of $\beta$. \citet{SuC11}, \citet{SuC12}, and \citet{CookS13} among others expanded the applications of this envelope method.

\citet{CookZ15} recently further expanded the envelope method to generalized linear regression models. The response belongs to an exponential family of the form~\eqref{eqn:exp-fam}. Let $\Ccal(\vartheta)=y\vartheta - b(\vartheta)$ and $W(\vartheta)=\Ccal^{''}(\vartheta)/E(\Ccal^{''}(\vartheta))$. They reparameterized $\vartheta(\alpha, \beta) = \alpha + \beta^T X$ to $\vartheta(\alpha, \beta)=a + \beta^T \{X - E(WX)\}$ with $a=\alpha + \beta^T E(WX)$ so that $a$ and $\beta$ are orthogonal. The asymptotic variance of $\betahat$ is then obtained as
$avar(\sqrt{n}\betahat)=\Vbf_{\beta \beta}(a, \beta)=\{E(-\Ccal^{''} E\{ [X-E(WX)][X-E(WX)]^T\} \}.$
With a reparameterization $\beta= \Gamma \eta$, the parameter $\eta$ is estimated using a Fisher scoring method fitting the GLM of $Y$ on $\Gamma^T X$. The partially maximized log-likelihood for $\Gamma$ is
$$L_n(\Gamma) = \sum_{i=1}^n \Ccal(\alpha + \etahat^T \Gamma^T X_i) - \frac{n}{2}\{\log|\Gamma^T \Srm_X \Gamma| + \log|\Gamma^T \Srm^{-1}_X\Gamma| + \log|\Srm_X|\},$$
where $\Srm_X$ is the sample covariance of $X$. This function is now optimized over a Grassmann manifold to obtain $\Gammahat$.

\section{Using ManifoldOptim}
\label{sec:usage}

We now describe use of the {ManifoldOptim} package. The core function is \texttt{manifold.optim}, which has the following interface.
\begin{verbatim}
manifold.optim(prob, mani.defn, method = "LRBFGS", x0 = NULL,
    has.hhr = FALSE, mani.params = get.manifold.params(),
    H0 = NULL, solver.params = get.solver.params())
\end{verbatim}
A typical call to \texttt{manifold.optim} involves four essential pieces: a problem \texttt{prob}, the choice of manifold \texttt{mani.defn}, configuration for the manifold \texttt{mani.params}, and configuration for the solver \texttt{solver.params}.

The \texttt{problem} encapsulates the objective function to be minimized, and, optionally, the gradient and Hessian functions. Analytical expressions of the gradient and Hessian are usually desirable but may be either tedious to compute or intractable. Numerical approximations to the gradient and Hessian via finite differences are used by default. There are three options for constructing a problem with {ManifoldOptim}: \texttt{RProblem}, \texttt{VectorManifoldOptimProblem}, and \texttt{MatrixManifoldOptimProblem}.

Generally, {ManifoldOptim} treats the optimization variable as a single vector, and the user must reshape the vector into the matrices and vectors specific to the problem at hand. This approach is similarly taken by the standard \texttt{optim} function in the {R} package {stats} \citep{Rco}.

The \texttt{method} argument specifies the algorithm to be used in the optimization; \texttt{LRBFGS} is the default method. A list of possible values for \texttt{method} is provided in Table~\ref{tab:algorithms}.

{ROPTLIB} provides solvers for at least nine commonly encountered manifolds at the time of this writing. In the first version of {ManifoldOptim}, we have focused on six manifolds: unconstrained Euclidean space, the Stiefel manifold, the Grassmann manifold, the unit sphere, the orthogonal group, and the manifold of symmetric positive definite matrices. As {ROPTLIB} continues development, and as the need arises in the \texttt{R} community, support for more solvers and manifolds will be added to {ManifoldOptim}.

The focus of the initial release {ManifoldOptim} has been on ease of use. {ROPTLIB} provides additional constructs to assist {C++} programmers in avoiding redundant computations and redundant memory usage, which can significantly improve run time of the optimization \citep{HAG2016}. This functionality is currently not exposed in {ManifoldOptim}, but may be considered in future versions.

\begin{table}
\centering
\small
\caption{List of optimization algorithms available in the \texttt{ManifoldOptim} package.}
\label{tab:algorithms}
\begin{tabular}{lll}
\hline
Code                      & Description \\
\hline
\texttt{RTRNewton}      & Riemannian trust-region Newton \citep{ABG07} \\
\texttt{RTRSR1}         & Riemannian trust-region symmetric rank-one update \citep{hag15}\\
\texttt{LRTRSR1}        & Limited-memory RTRSR1 \citep{hag15}\\
\texttt{RTRSD}          & Riemannian trust-region steepest descent \citep{AMS08}\\
\texttt{RNewton}        & Riemannian line-search Newton \citep{AMS08}\\
\texttt{RBroydenFamily} & Riemannian Broyden family \citep{HGA15} \\
\texttt{RWRBFGS} and \texttt{RBFGS} & Riemannian BFGS \citep{RW12, HGA15}\\
\texttt{LRBFGS}         & Limited-memory RBFGS \citep{HGA15}\\
\texttt{RCG}            & Riemannian conjugate gradients \citep{AMS08, si13}\\ 
\texttt{RSD}            & Riemannian steepest descent \citep{AMS08}\\
\hline
\end{tabular}
\end{table}

\subsection{Solving the Brockett Problem}
We revisit the Brockett problem described in section~\ref{sec:alg} to illustrate the use of {ManifoldOptim}. The gradient for the objective function can be obtained in closed form as
\begin{align*}
\nabla f(X) = 2 B X D.
\end{align*}
Denoting $\otimes$ as the Kronecker product operator, we may write $\vecc(\nabla f(X)) = 2 (D \otimes B) \vecc(X)$ to obtain the Hessian $\nabla^2 f(\vecc(X)) = 2 (D \otimes B)$. It should be noted that the usual Euclidean gradient and Hessian are to be computed here, ignoring manifold constraints. The {ROPTLIB} library can also work with the Riemannian gradient, but this is not exposed in {ManifoldOptim}.

We set the matrices $B$ and $D$ to prepare a concrete instance of the problem.
\begin{verbatim}
set.seed(1234)
n <- 150; p <- 5
B <- matrix(rnorm(n*n), nrow=n)
B <- B + t(B)    # Make symmetric
D <- diag(p:1, p)
\end{verbatim}
The objective and gradient are coded in {R} as follows.
\begin{verbatim}
tx <- function(x) { matrix(x, n, p) }
f <- function(x) { X <- tx(x); Trace( t(X) %*% B %*% X %*% D ) }
g <- function(x) { X <- tx(x); 2 * B %*% X %*% D }
\end{verbatim}
The \texttt{tx} function reshapes an $np$-dimensional point from the solver into an $n \times p$ matrix, which can then be evaluated by the \texttt{f} and \texttt{g} functions. We will initially select a solver that does not make use of a user-defined Hessian function.

We first consider an \texttt{RProblem}, where all functions are programmed in {R}. This provides a convenient way to code a problem, but may face some performance limitations. In general, functions coded in {R} tend to be much slower than similar functions written in {C++}, especially if they cannot be vectorized. With an \texttt{RProblem}, each evaluation of the objective, gradient, and Hessian by the solver incurs the overhead of a call from {C++} to a function defined in {R}. Having noted the potential performance issues, we now proceed with construction of an \texttt{RProblem}.
\begin{verbatim}
mod <- Module("ManifoldOptim_module", PACKAGE = "ManifoldOptim")
prob <- new(mod$RProblem, f, g)
\end{verbatim}

The \texttt{Module} construct \citep{EddelbuettelFrancois2016} is useful for the internal implementation of {ManifoldOptim}. When constructing a module in {R}, the user implicitly creates a {C++} problem which can be accessed by the {C++} solver.

The Brockett function is to be optimized over a Stiefel manifold, so we use \texttt{get.stiefel.defn} to create the specification. Additionally, we set software parameters for the manifold and solver via the \texttt{get.manifold.params} and \texttt{get.solver.params} functions.
\begin{verbatim}
mani.defn <- get.stiefel.defn(n, p)
mani.params <- get.manifold.params(IsCheckParams = TRUE)
solver.params <- get.solver.params(DEBUG = 0, Tolerance = 1e-4,
    Max_Iteration = 1000, IsCheckParams = TRUE)
\end{verbatim}

The argument \texttt{IsCheckParams = TRUE} requests useful information to be printed on the console for either the manifold or solver, depending where it is placed. \texttt{DEBUG} is an integer that sets the verbosity of the solver during iterations; The lowest verbosity is \texttt{DEBUG = 0}, which prints no debugging information. \texttt{Max\_Iteration} and \texttt{Tolerance} set the maximum iteration and tolerance to determine when the solver will halt. More extensive descriptions for the arguments are given in the {ManifoldOptim} manual.

A starting value for the optimization can be specified by the argument \texttt{x0} in the call of the function \texttt{manifold.optim}. If available, a good initial value can assist the solver by reducing the time to find a solution and improving the quality of the solution when multiple local optima are present. If no initial value is given, the optimizer will select one at random from the given manifold. In this case of the Brockett problem, we consider the following initial value.
\begin{verbatim}
x0 <- as.numeric(diag(n)[,1:p])
\end{verbatim}
Now that we have specified the problem, manifold definition, software parameters for the manifold and solver, an algorithm, and an initial value, we can invoke the optimizer through the \texttt{manifold.optim} function.
\begin{verbatim}
res <- manifold.optim(prob, mani.defn, x0 = x0, method = "RTRSR1",
    mani.params = mani.params, solver.params = solver.params)
\end{verbatim}
Upon completion, \texttt{manifold.optim} produces a result, \texttt{res}, which contains the quantities listed below.
\begin{verbatim}
> names(res)
 [1] "xopt"          "fval"          "normgf"        "normgfgf0"
 [5] "iter"          "num.obj.eval"  "num.grad.eval" "nR"
 [9] "nV"            "nVp"           "nH"            "elapsed"
[13] "funSeries"     "gradSeries"    "timeSeries"
\end{verbatim}

The {ManifoldOptim} manual gives a description of each of these items. The most important output is the solution \texttt{xopt}, which optimizes the objective function. Using \texttt{tx} to reshape it into a matrix, its first rows are given below.
\begin{verbatim}
> head(tx(round(res$xopt, digits=4)))
        [,1]    [,2]    [,3]    [,4]    [,5]
[1,]  0.1688 -0.1100 -0.0876 -0.0396 -0.0520
[2,] -0.0075  0.1275 -0.1243  0.0486  0.1053
[3,]  0.0956 -0.0249  0.0396 -0.0008 -0.0134
[4,] -0.0381 -0.1710 -0.0477 -0.0142  0.0660
[5,] -0.1239 -0.0207  0.1199 -0.1236 -0.1112
[6,] -0.0296 -0.0049  0.0511 -0.1352 -0.0129
\end{verbatim}

Solvers such as \texttt{RNewton} \citep{AMS08} require a Hessian function to be specified for the problem. If a Hessian function is required by the solver but not provided by the user, a numerical approximation will be used. This provides a convenient default, but can be slow and potentially inaccurate. We briefly illustrate the use of a Hessian function for the Brockett \texttt{RProblem}.

Treating the optimization variable $\vecc(X)$ as a $q$-dimensional vector, the $q \times q$ Hessian function $\nabla^2 f(\vecc(X))$ is programmed by coding the action $[\nabla^2 f(\vecc(X))] \eta$ for a given $\eta$ in the tangent space to the manifold at $X$. Further detail on the implementation of this action for either line search or trust region solver algorithms is given in \citep{HAG2016} and \citep{AMS08}. For the Brockett problem, this becomes the matrix multiplication $2 (D \otimes B) \eta$. For an \texttt{RProblem}, this expression can be specified as the third argument of the constructor.

%
\begin{verbatim}
h <- function(x, eta) { 2 * (D %x% B) %*% eta }
prob <- new(mod$RProblem, f, g, h)
\end{verbatim}
We note that the efficiency of the \texttt{h} function can be greatly improved by avoiding direct computation of $D \otimes B$, but proceed with this simple coding to facilitate the demonstration.

{ROPTLIB} provides a diagnostic to check correctness of the gradient and Hessian. It produces two outputs: the first uses the starting value, and the second uses the solution obtained from the optimizer. If the quantity \texttt{(fy-fx)/<gfx,eta>} is approximately 1 for some interval within the first output, it is an indication that the gradient function is correct. If the quantity \texttt{(fy-fx-<gfx,eta>)/<0.5 eta, Hessian eta>} is approximately 1 for some interval within the second output, it indicates that the Hessian function is correct. See the {ROPTLIB} manual for details. The diagnostic is requested by setting the \texttt{IsCheckGradHess} option  in the solver.
\begin{verbatim}
solver.params <- get.solver.params(IsCheckGradHess = TRUE)
res <- manifold.optim(prob, mani.defn, method = "RNewton",
    mani.params = mani.params, solver.params = solver.params, x0 = x0)
\end{verbatim}

The first output validates the gradient function.
\begin{scriptsize}
\begin{verbatim}
i:0,|eta|:1.000e+02,(fy-fx)/<gfx,eta>:-1.463e-04,(fy-fx-<gfx,eta>)/<0.5 eta, Hessian eta>:-2.524e+01
...
i:28,|eta|:3.725e-07,(fy-fx)/<gfx,eta>:1.000e+00,(fy-fx-<gfx,eta>)/<0.5 eta, Hessian eta>:-4.147e-01
i:29,|eta|:1.863e-07,(fy-fx)/<gfx,eta>:1.000e+00,(fy-fx-<gfx,eta>)/<0.5 eta, Hessian eta>:2.441e-01
i:30,|eta|:9.313e-08,(fy-fx)/<gfx,eta>:1.000e+00,(fy-fx-<gfx,eta>)/<0.5 eta, Hessian eta>:-1.669e+01
...
i:34,|eta|:5.821e-09,(fy-fx)/<gfx,eta>:1.000e+00,(fy-fx-<gfx,eta>)/<0.5 eta, Hessian eta>:5.574e+02
\end{verbatim}
\end{scriptsize}
The second output validates the Hessian function.
\begin{scriptsize}
\begin{verbatim}
i:0,|eta|:1.000e+02,(fy-fx)/<gfx,eta>:1.556e+03,(fy-fx-<gfx,eta>)/<0.5 eta, Hessian eta>:3.823e-04
...
i:21,|eta|:4.768e-05,(fy-fx)/<gfx,eta>:2.939e+00,(fy-fx-<gfx,eta>)/<0.5 eta, Hessian eta>:1.000e+00
i:22,|eta|:2.384e-05,(fy-fx)/<gfx,eta>:1.970e+00,(fy-fx-<gfx,eta>)/<0.5 eta, Hessian eta>:1.000e+00
i:23,|eta|:1.192e-05,(fy-fx)/<gfx,eta>:1.485e+00,(fy-fx-<gfx,eta>)/<0.5 eta, Hessian eta>:1.000e+00
...
i:34,|eta|:5.821e-09,(fy-fx)/<gfx,eta>:1.008e+00,(fy-fx-<gfx,eta>)/<0.5 eta, Hessian eta>:3.368e+01
\end{verbatim}
\end{scriptsize}

\subsection{Coding a Problem in C++}
\label{sec:vector-problem}

For optimization problems which require intensive computation or will be evaluated repeatedly, it may be worth investing some additional effort to write the problem in {C++}. We now illustrate by converting the Brockett problem to {C++}. To proceed, we extend the \texttt{VectorManifoldOptimProblem} class within {ManifoldOptim}. A \texttt{VectorManifoldOptimProblem} has objective, gradient, and Hessian functions coded in {C++} using {RcppArmadillo}. The user implements this problem type by extending the abstract \texttt{VectorManifoldOptimProblem} class.

The class \texttt{BrockettProblem} can be written as follows.
\begin{verbatim}
Rcpp::sourceCpp(code = '
//[[Rcpp::depends(RcppArmadillo,ManifoldOptim)]]
#include <RcppArmadillo.h>
#include <ManifoldOptim.h>

using namespace Rcpp;
using namespace arma;

class BrockettProblem : public VectorManifoldOptimProblem
{
public:
    BrockettProblem(const arma::mat& B, const arma::mat& D)
    : VectorManifoldOptimProblem(false,false), _B(B), _D(D) { }

    virtual ~BrockettProblem() { }

    double objFun(const arma::vec& x) const {
        arma::mat X;
        tx(X, x);
        return arma::trace(X.t() * _B * X * _D);
    }

    arma::mat gradFun(const arma::vec& x) const {
        arma::mat X;
        tx(X, x);
        return 2 * _B * X * _D;
    }

    arma::vec hessEtaFun(const arma::vec& x, const arma::vec& eta) const {
        return 2 * arma::kron(_D, _B) * eta;
    }

    void tx(arma::mat& X, const arma::vec& x) const {
        X = x;
        X.reshape(_B.n_rows, _D.n_rows);
    }

    const arma::mat& GetB() const { return _B; }
    const arma::mat& GetD() const { return _D; }

private:
    arma::mat _B;
    arma::mat _D;
};

RCPP_MODULE(Brockett_module) {
    class_<BrockettProblem>("BrockettProblem")
    .constructor<mat,mat>()
    .method("objFun", &BrockettProblem::objFun)
    .method("gradFun", &BrockettProblem::gradFun)
    .method("hessEtaFun", &BrockettProblem::hessEtaFun)
    .method("GetB", &BrockettProblem::GetB)
    .method("GetD", &BrockettProblem::GetD)
    ;
}
')
\end{verbatim}
Note that we have used \texttt{sourceCpp} to compile the source code as a string from within {R}. It is also possible to write the {C++} code in a standalone file and compile it using \texttt{sourceCpp}, or to have it compiled within the build of your own custom package; refer to the {Rcpp} documentation for more details. In addition to defining the \texttt{BrockettProblem} class, our source code defines a module which allows \texttt{BrockettProblem} objects to be constructed and manipulated from {R}. Specifically, the constructor, objective, gradient, Hessian, and accessor functions for $B$ and $D$ can be called from {R}.

The \texttt{tx} function is responsible for reshaping the vector \texttt{x} into variables which can be used to evaluate the objective, gradient, and Hessian functions. In this case, the optimization variable is a single matrix, so we simply reshape vector \texttt{x} into the matrix \texttt{X} using the {Armadillo} \texttt{reshape} function. If \texttt{x} were to contain multiple variables, the \texttt{tx} function could be modified to have multiple output arguments in its definition.

To define a \texttt{BrockettProblem} we must specify the $B$ and $D$ matrices via the constructor. Using the same $B$ and $D$ as in the previous section, we can invoke the constructor through {R}.
\begin{verbatim}
prob <- new(BrockettProblem, B, D)
\end{verbatim}
Setting the manifold definition, manifold configuration, and solver configuration is done in the same way as before.
\begin{verbatim}
mani.defn <- get.stiefel.defn(n, p)
mani.params <- get.manifold.params(IsCheckParams = TRUE)
solver.params <- get.solver.params(DEBUG = 0, Tolerance = 1e-4,
    Max_Iteration = 1000, IsCheckParams = TRUE)
\end{verbatim}
We use the same initial value as before. Thanks to our {Rcpp} module, we can call the problem functions from {R}, which makes them easier to test.
\begin{verbatim}
x0 <- as.numeric(diag(n)[,1:p])
eta <- rnorm(n*p)
prob$objFun(x0)
prob$gradFun(x0)
prob$hessEtaFun(x0, eta)
\end{verbatim}
We may now invoke the solver.
\begin{verbatim}
res <- manifold.optim(prob, mani.defn, method = "RTRSR1",
    mani.params = mani.params, solver.params = solver.params, x0 = x0)
\end{verbatim}
After obtaining a solution, our first task will likely be to reshape it from a vector to a matrix.
\begin{verbatim}
tx <- function(x) { matrix(x, n, p) }
head(tx(res$xopt))
\end{verbatim}
The result is close to the one obtained in the previous section and is therefore not shown here.

\subsection{An Extra C++ Convenience for Functions of a Single Matrix}
\label{sec:matrix-problem}
For problems where the optimization variable is a single matrix (e.g.~optimization on a Grassmann manifold), users may extend the class \texttt{MatrixManifoldOptimProblem}. Doing this is very similar to extending \texttt{VectorManifoldOptimProblem}, but gives the convenience of having the optimization variable presented to problem functions as a matrix, partially avoiding the need for reshaping code.

Continuing with the Brockett example, the \texttt{MatrixManifoldOptimProblem} version is written as follows.
\begin{verbatim}
Rcpp::sourceCpp(code = '
//[[Rcpp::depends(RcppArmadillo,ManifoldOptim)]]
#include <RcppArmadillo.h>
#include <ManifoldOptim.h>

using namespace Rcpp;
using namespace arma;

class BrockettProblem : public MatrixManifoldOptimProblem
{
public:
    BrockettProblem(const arma::mat& B, const arma::mat& D)
    : MatrixManifoldOptimProblem(false), _B(B), _D(D) { }

    virtual ~BrockettProblem() { }

    double objFun(const arma::mat& X) const {
        return arma::trace(X.t() * _B * X * _D);
    }

    arma::mat gradFun(const arma::mat& X) const {
        return 2 * _B * X * _D;
    }

    arma::vec hessEtaFun(const arma::mat& X, const arma::vec& eta) const {
        return 2 * arma::kron(_D, _B) * eta;
    }

    const arma::mat& GetB() const { return _B; }
    const arma::mat& GetD() const { return _D; }

private:
    arma::mat _B;
    arma::mat _D;
};

RCPP_MODULE(Brockett_module) {
    class_<BrockettProblem>("BrockettProblem")
    .constructor<mat,mat>()
    .method("objFun", &BrockettProblem::objFun)
    .method("gradFun", &BrockettProblem::gradFun)
    .method("hessEtaFun", &BrockettProblem::hessEtaFun)
    .method("GetB", &BrockettProblem::GetB)
    .method("GetD", &BrockettProblem::GetD)
    ;
}
')
\end{verbatim}
The code is very similar to Section~\ref{sec:matrix-problem}, except now the \texttt{tx} function is no longer necessary. Note that the solver still treats the optimization variable as a vector, and the initial value \texttt{x0} is coded as a vector, but the objective, gradient, and Hessian expect a matrix. To evaluate the problem functions from {R} at \texttt{x0}, we can transform \texttt{x0} to a matrix using our \texttt{tx} function.
\begin{verbatim}
x0 <- as.numeric(diag(n)[,1:p])
eta <- rnorm(n*p)
tx <- function(x) { matrix(x, n, p) }
prob$objFun(tx(x0))
prob$gradFun(tx(x0))
prob$hessEtaFun(tx(x0), eta)
\end{verbatim}

\subsection{Product Manifold}
\label{sec:product-manifold}
A product of manifolds can be used to optimize jointly over multiple optimization variables. As a demonstration, consider observing a random sample $Y_1, \ldots, Y_n$ from a $p$-variate normal distribution with mean $\mu$ constrained to the unit sphere and symmetric positive definite variance $\Sigma$. The maximum likelihood estimators are obtained as
\begin{align*}
(\muhat, \Sigmahat) = \argmax_{\mu \in \Scal^p, \Sigma\in \Scal_{p}^+} \left\{
-\frac{np}{2} \log(\pi)- \frac{n}{2} \log|\Sigma| -\frac{1}{2} \sum_{i=1}^n (Y_i - \mu)^T \Sigma^{-1} (Y_i - \mu)
\right\}.
\end{align*}
To code this problem, let us first generate an example dataset.
\begin{verbatim}
set.seed(1234)
n <- 400
p <- 3
mu.true <- rep(1/sqrt(p), p)
Sigma.true <- diag(2,p) + 0.1
y <- mvtnorm::rmvnorm(n, mean = mu.true, sigma = Sigma.true)
\end{verbatim}
Next we define the objective, and a function \texttt{tx} which reshapes a dimension $p + p^2$ vector into a $p$-dimensional vector and a $p \times p$ matrix.
\begin{verbatim}
tx <- function(x) {
    list(mu = x[1:p], Sigma = matrix(x[1:p^2 + p], p, p))
}
f <- function(x) {
    par <- tx(x)
    -sum(mvtnorm::dmvnorm(y, mean = par$mu, sigma = par$Sigma, log = TRUE))
}
mod <- Module("ManifoldOptim_module", PACKAGE = "ManifoldOptim")
prob <- new(mod$RProblem, f)
\end{verbatim}

\noindent We have used an \texttt{RProblem} so that the objective function can be specified as an {R} function. The negative of the log-likelihood has been given as the objective to achieve a maximization. We have not specified gradient or Hessian functions so that numerical approximations will be used by the solver. A product manifold definition is now constructed for the problem using unit sphere manifold $\Scal^p$ and SPD manifold $\Scal_{p}^+$ definitions. We also specify options for the manifold and solver.
\begin{verbatim}
mani.defn <- get.product.defn(get.sphere.defn(p), get.spd.defn(p))
mani.params <- get.manifold.params()
solver.params <- get.solver.params(Tolerance = 1e-4)
\end{verbatim}
Note that, for any unconstrained parameters, the \texttt{get.euclidean.defn} function can be used in the product of manifolds. We now give an initial value and invoke the solver.
\begin{verbatim}
mu0 <- diag(1, p)[,1]
Sigma0 <- diag(1, p)
x0 <- c(mu0, as.numeric(Sigma0))
res <- manifold.optim(prob, mani.defn, method = "LRBFGS",
    mani.params = mani.params, solver.params = solver.params, x0 = x0)
\end{verbatim}
The following result is produced.
\begin{verbatim}
> tx(res$xopt)
$mu
[1] 0.5076927 0.7055063 0.4944785

$Sigma
           [,1]      [,2]       [,3]
[1,]  2.0012674 0.1353931 -0.1739273
[2,]  0.1353931 2.0275420  0.2641577
[3,] -0.1739273 0.2641577  2.0936022
\end{verbatim}
The result for $\Sigma$ may not be exactly symmetric due to numerical error. If exact symmetry is needed for $\Sigma$, the symmetric part can be extracted as the estimate.
\begin{verbatim}
S <- tx(res$xopt)$Sigma
Sigma.hat <- 1/2 * (S + t(S))
\end{verbatim}

\subsection{Customizing Line Search}
\label{sec:line-search}
Line search algorithms are analogous to the method of gradient descent. As described in detail in \citep{AMS08}, they are based on an update
\begin{equation}
x_{k+1} = x_k + t_{k} \eta_{k},
\end{equation}
where $\eta_k \in \Rbb^n$ is the search direction and $t_k \in \Rbb$ is the step size. Determining a step size on the curved surface of a manifold requires special consideration beyond what is needed for Euclidean space. The concept of a retraction mapping is employed to move in the direction of a tangent vector while staying on the manifold. Several line search algorithms are available in the {ROPTLIB} library, along with options to customize the search. Some of these options can be configured through the {R} interface.

As an example, consider the product manifold example from Section~\ref{sec:product-manifold}. Several line search parameters are specified below. In particular, \texttt{LineSearch\_LS = 1} corresponds to the Wolfe line search method.
\begin{verbatim}
solver.params <- get.solver.params(Tolerance = 1e-4, IsCheckParams = TRUE,
    LineSearch_LS = 1, Accuracy = 2e-4, Initstepsize = 1/2)
\end{verbatim}
Now, running the LRBFGS optimizer results in the following output.
\begin{verbatim}
> res <- manifold.optim(prob, mani.defn, method = "LRBFGS",
    mani.params = mani.params, solver.params = solver.params, x0 = x0)
GENERAL PARAMETERS:
Stop_Criterion:       GRAD_F_0[YES],	Tolerance     :         0.0001[YES]
Max_Iteration :           1000[YES],	Min_Iteration :              0[YES]
OutputGap     :              1[YES],	DEBUG         :       NOOUTPUT[YES]
LINE SEARCH TYPE METHODS PARAMETERS:
LineSearch_LS :          WOLFE[YES],	LS_alpha      :         0.0001[YES]
LS_beta       :          0.999[YES],	Initstepsize  :            0.5[YES]
Minstepsize   :    2.22045e-16[YES],	Maxstepsize   :           1000[YES]
Accuracy      :         0.0002[YES],	Finalstepsize :              1[YES]
Num_pre_funs  :              0[YES],	InitSteptype  :     QUADINTMOD[YES]
LRBFGS METHOD PARAMETERS:
nu            :         0.0001[YES],	mu            :              1[YES]
isconvex      :              0[YES],	LengthSY      :              4[YES]
\end{verbatim}
Observe that the selected line search options are reported in the output. The full list of available line search options can be found in \citet{HAG2016}. At this time, a user-defined line search algorithm (\texttt{LineSearch\_LS = 5}) cannot be specified via {R}, but support will be considered for a future release of {ManifoldOptim}.

\section{Examples with Statistical Methods}
\label{sec:examples}

This section demonstrates the use of {ManifoldOptim} in the three main methodologies presented in section~\ref{sec:alg}: MADE, PFC, and envelope regression.

\subsection{Minimum Average Deviance Estimation}
\label{sec:made-study}

We consider an application of MADE, which was summarized in section~\ref{sec:made}, on a fishing dataset. The data are adapted from \citet{Bailey09}, who were interested in how certain deep-sea fish populations were impacted when commercial fishing began in locations with deeper water than in previous years. The dataset has 147 observations and three continuous predictors: \texttt{density}, \texttt{log(meandepth)}, and \texttt{log(sweptarea)} which are, respectively, foliage density index, the natural logarithm of mean water depth per site, and the natural logarithm of the adjusted area of the site. The response \texttt{totabund} is the total number of fish counted per site. Our goal is to find a sufficient reduction of the three predictors to capture all regression information between the response and the predictors. To apply MADE to the fishing data, we assume a Poisson family for \eqref{eqn:exp-fam}.

A full MADE program is included in the supplemental material. Step 2(b) of the iterative algorithm from section~\ref{sec:made} is a Stiefel manifold optimization and can be carried out with {ManifoldOptim}. We make use of the \texttt{RCG} solver along with code for the objective function and its derivative with respect to $B$; these codes are based on the overall MADE objective \eqref{eqn:made-glm-obj}.

\begin{figure}[!h]
\centering
\begin{tabular}{cc}

\includegraphics[width=0.4\textwidth]{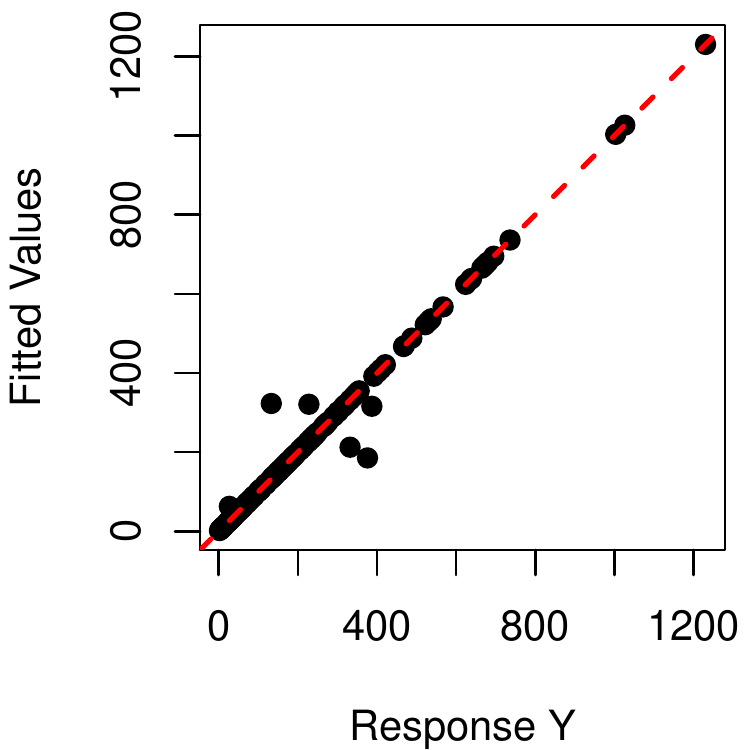} &
\includegraphics[width=0.4\textwidth]{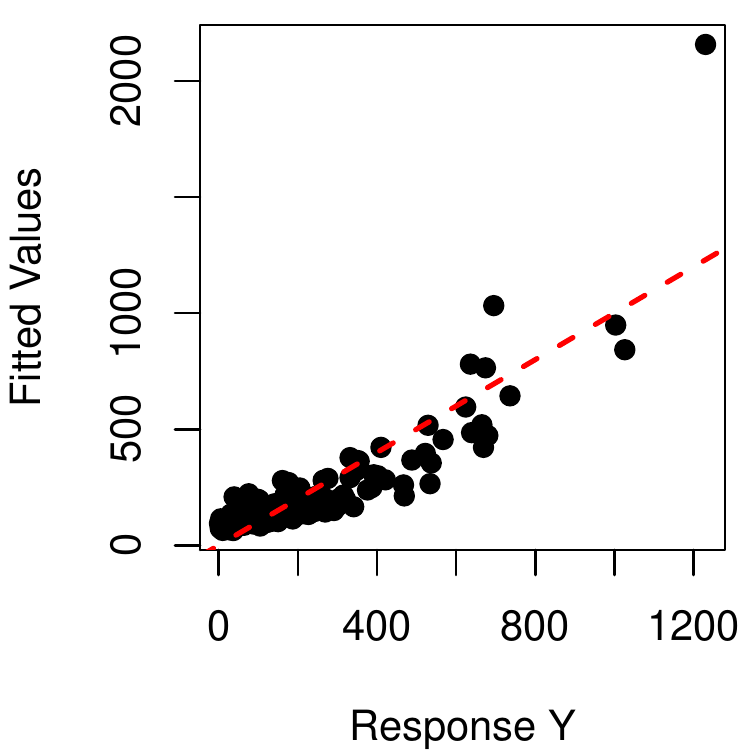} \\
(a) MADE & (b) GLM

\end{tabular}
\caption{Plots of the fitted response $\Yhat$ against the observed $Y$.}
\label{fig:fishing}
\end{figure}

The observed response is plotted against the fitted MADE response with $d=1$ in Figure~\ref{fig:fishing}(a), and using the Poisson generalized linear regression model in Figure~\ref{fig:fishing}(b). Ideally, these points would follow a 0-intercept unit-slope line (plotted as the dashed line). The observation with $y = 1230$ appears to be an outlier under the estimated model.

\subsection{Principal Fitted Components}
\label{sec:pfc-study}

We now consider a simulation study using the PFC model from section~\ref{sec:pfc} to evaluate the performance of {ManifoldOptim}. The datasets were generated using $n=300$ observations with response $\Ybb \in \Rbb^{n}$ from the normal distribution with mean 0 and variance 9. The matrix of basis functions $\Fbb \in \Rbb^{n \times 2}$ was obtained as $(\Ybb, |\Ybb|)$ and is column-wise centered to have sample mean 0. We obtained $\Gamma$ as two eigenvectors of a $p \times p$ generated positive definite matrix. The data matrix of the predictors $\Xbb \in \Rbb^{n \times p}$ was obtained as $\Xbb = \Fbb \Gamma^T + \Ebb$, with error term $\Ebb \in \Rbb^{n \times p}$ generated from the multivariate normal distribution with mean 0 and variance $\Delta$.



Two structures were used for the covariance $\Delta$. The first was an unstructured covariance matrix $\Delta_{\urm} = U^TU$ where $U$ is a $p \times p$ matrix with entries from the standard normal distribution. The second was an envelope structure $\Delta_{\erm}= \Gamma \Omega \Gamma^T + \Gamma_0 \Omega_0 \Gamma_0$ with $\Omega \in \Rbb^{2 \times 2}$ and $\Omega_0 \in \Rbb^{(p-2) \times (p-2)}$. The value of $\Gamma$ is obtained from the first two eigenvectors of the unstructured $\Delta$. The remaining eigenvectors are used for the value of $\Gamma_0$. These two structures correspond to the two models discussed in section~\ref{sec:pfc}. The likelihood~\eqref{unst} corresponds to the unstructured $\Delta_{\urm}$ while the likelihood function \eqref{env} corresponds to the envelope structure $\Delta_{\erm}$. The parameters are then $(\Gamma, \Delta)$ in the unstructured PFC case, and $(\Gamma, \Omega, \Omega_0)$ for the envelope PFC. We note that in the case of the envelope structure, $\Gamma_0$ is the orthogonal completion of $\Gamma$ so that $\Gamma \Gamma^T + \Gamma_0 \Gamma_0^T =I$.

Estimation methods for these two PFC models exist. \citet{c+f+09} provide closed-form solutions for estimation of $\Gamma$ and $\Delta$ in the unstructured case. For the envelope PFC, \citet{Cook07} provided a maximum likelihood estimation over a Grassmann manifold for $\Gamma$ while $\Omega$ and $\Omega_0$ have a closed-form that depends on $\Gamma$ and $\Gamma_0$. The goal here is to evaluate the optimization and compare its performance to the closed-form solutions. Moreover, a product manifold is used so that the estimation of $(\Gamma, \Delta)$ is done once instead of a typical alternating procedure that alternates between holding certain values constant and optimizing over them.

To proceed, we coded the objective and gradients functions under both setups. For each dataset generated in the simulation, the closed-form solutions were obtained and compared to the solutions obtained via {ManifoldOptim}. We compared the true parameter $\Gamma$ to the estimate $\Gammahat$ using the subspace distance $\rho(\Gamma, \Gammahat)=\|(I-\Gammahat \Gammahat^T)\Gamma\|$ suggested by \citet{Xia02}. The true and estimated covariance were compared using $d(\Delta, \Deltahat)=\tr\{(\Delta - \Deltahat)^T(\Delta - \Deltahat)\}$. One hundred replications were used for the simulation. The estimated mean distances $d(\Delta, \Deltahat) $ and $\rho(\Gamma, \Gammahat)$ are reported in Tables~\ref{tab:evals_unstruct} and~\ref{tab:evals_envelope} with their standard errors in parentheses.

The results indicate that both forms of optimization perform comparably. Thus, the product manifold in {ManifoldOptim} provides an alternative form of optimization which can ease implementation. Also note that the difference between the true and estimated covariance is significantly less for the envelope-$\Delta$ PFC, since there are less parameters to estimate in this case. Optimization was carried out using $\Sigmahat$ and $\Sigmahat_{\res}$ as initial values for $\Delta$; the results were similar to the classical closed form method.
The full code is provided as a supplement to this article.
\begin{table}[h!]
\centering
\caption{Comparison of classical estimation and manifold optimization using {ManifoldOptim} of unstructured-$\Delta$ PFC parameters.}
\label{tab:evals_unstruct}
\begin{tabular}{l||c|c}
Algorithm     & $d(\Delta, \Deltahat)$ & $\rho(\Gamma, \Gammahat)$ \\
\hline
Classical     & 2.67 (0.125)           & 0.17 (0.004) \\
ManifoldOptim & 2.68 (0.125)           & 0.17 (0.004) \\
\end{tabular}
\end{table}
\begin{table}[h!]
\centering
\caption{Comparison of classical estimation and manifold optimization using {ManifoldOptim} of envelope-$\Delta$ PFC parameters.}
\label{tab:evals_envelope}
\begin{tabular}{l||r|r|r}
Algorithm     & $d(\Omega, \Omegahat)$ & $d(\Omega_0, \Omegahat_0)$ & $\rho(\Gamma, \Gammahat)$ \\
\hline
Classical     & 0.03 (0.003)           & 0.36 (0.008)               & 0.28 (0.005)\\
ManifoldOptim & 0.03 (0.003)           & 0.35 (0.008)               & 0.28 (0.005)
\end{tabular}
\end{table}

\subsection{Envelope Models}
\label{sec:env-study}

We consider a simulation in \citep{CookZ15} to compare the results using {ManifoldOptim} to results using the three algorithms in \citet{CookZ15} during a logistic regression. We generated 150 independent observations using $Y_i \mid X_i \sim \text{Bernoulli}(\logit^{-1}(\beta^T \Xrm_i))$, with $\beta=(\beta_1,\beta_2)^T=(0.25, 0.25)^T$ and $\Xrm_i \sim N(0, \Sigma_{\Xrm})$. We let $v_{1}=\spn(\beta)$ be the principal eigenvector of $\Sigma_{\Xrm}$ with eigenvalue 10 and let the second eigenvector $v_{2}$ be the orthogonal completion with eigenvalue of 0.1 such that $v_{1}v_{1}^T + v_{2}v_{2}^T = I$. This example is compelling because of the collinearity during construction of $\Sigma_{\Xrm}$. Consequently, this causes poor estimates when using a standard generalized linear model (GLM).

%
\begin{table}[h]
\centering
\caption{Comparison of estimators between truth, Cook's algorithm 1 and product manifold using {ManifoldOptim}.}
\label{tab:table4}
\begin{tabular}{l||r|r}
Algorithm                             & $\beta_1$ & $\beta_2$ \\
\hline
Truth                                 &  0.25     & 0.25 \\
GLM                                   & -0.12     & 0.63 \\
Cook's Algorithm 1 with ManifoldOptim &  0.25     & 0.24 \\
Product Manifold Optimization         &  0.25     & 0.24 \\
\end{tabular}
\end{table}
Three algorithms are considered in this section, namely GLM, Algorithm 1 in \citet{CookZ15}, and a product manifold optimization. Both Algorithm 1 and the product manifold are constructed using functionality in {ManifoldOptim}. In the case of Algorithm 1, a Grassmann optimization is performed along with Fisher scoring in an iterative fashion. During the product manifold optimization the Grassmann manifold is optimized for $\Gamma$; $\eta$ and $\alpha$ are estimated over the Euclidean manifold. Numerical derivatives are calculated with {ManifoldOptim} instead of using the analytical expression from \citet{CookZ15} since the latter pertains only to Grassmann manifold optimization. In all cases the LRBFGS \citep{HGA15} algorithm is used for optimizing over manifolds with the objective function given by equation 3.7 in \citet{CookZ15}. To avoid poor local maxima, the optimization was repeated several times from random starting points. Out of the many estimates at convergence, the one with the largest likelihood value was selected. The full code is provided as a supplement to this article.

By examining the results shown in Table~\ref{tab:table4} it is immediately clear that the collinearity causes serious issues with classical methods of estimation such as GLM. The estimates in this case are not even close to the true values. However, both of the other methods perform well. Similar to Section \ref{sec:pfc-study}, also note that the alternating optimization method shown in Cook's Algorithm 1 provides comparable results to the product manifold generated using {ManifoldOptim}. Since the product manifold is easier to implement, it can be a viable alternative.



\section{Conclusions}
\label{sec:conclusions}
We have presented {ManifoldOptim}, an {R} package for optimization on Riemannian Manifolds. {ManifoldOptim} fills a need for manifold optimization in {R} by making the functionality of {ROPTLIB} accessible to {R} users. The package so far deals with optimization on the Stiefel manifold, Grassmann manifold, unit sphere manifold, the manifold of positive definite matrices, the manifold of low rank matrices, the orthogonal group, and Euclidean space. We discussed basic usage of the package and demonstrated coding of problems in plain {R} or with {C++}. Furthermore, we demonstrated the product space of manifolds to optimize over multiple variables where each is constrained to its own manifold. Optimization over manifold-constrained parameters is needed for emerging statistical methodology in areas such as sufficient dimension reduction and envelope models. We have discussed several applications of such models, and demonstrated ways in which {ManifoldOptim} could practically be applied. We hope that availability of this package will facilitate exploration of these methodologies by statisticians and the {R} community.


\end{document}